\documentclass[11pt]{article}  
\usepackage{graphicx}
\usepackage{amssymb,amsmath}
\newcommand{\be}{\begin{equation}}
\newcommand{\ee}{\end{equation}}
\newcommand{\bea}{\begin{eqnarray}}
\newcommand{\eea}{\end{eqnarray}}
\begin{document}
\begin{center} 
{\LARGE{\bf Quantum geometric potential induced  conformational transitions in elastic helical nanoribbons}}
\end{center} 
\vspace{.3in}
\begin{center} 
{{\bf Radha Balakrishnan$^{(1)}$, Rossen Dandoloff$^{(2)}$, and Avadh Saxena$^{(3)}$}}\\ 
{$^{(1)}$The Institute of Mathematical Sciences, Chennai 600 113, India \\
$^{(2)}$Department of Condensed Matter Physics and Microelectronics, Faculty of Physics, 
Sofia University, 5 Blvd. J. Bourchier, 1164 Sofia, Bulgaria \\ 
$^{(3)}$Theoretical Division and Center for Nonlinear Studies, Los
Alamos National Laboratory, Los Alamos, New Mexico 87545, USA}
\end{center} 

\vspace{.6in}
{\bf {Abstract:}}

We consider
an {\em elastic} helical nanoribbon that can take on various conformations, and
 study the effect of placing  a quantum particle such as an electron  on its  curved surface. Using a  modified Canham-Helfrich model  for the elastic energy, we write down the local elastic potential for the ribbon in terms of its bending rigidity, mean curvature $M$ and Gaussian curvature $K$. As is well known, the Schr\"odinger equation of a particle confined to a {\em rigid}  curved surface is found using da Costa's formulation.
  It has a  purely  quantum geometric potential  which  depends on $M$ and $K$.   The Schr\"odinger equation of a particle  on an {\em elastic } curved surface will therefore have a  total potential  comprising  quantum and elastic potentials. Since both potentials depend on $M$ and $K$,  an intricate interplay  arises between them. We compute $M$ and $K$ for a helical ribbon and derive the total potential which depends on the conformation and is thus geometric in nature. Defining  a dimensionless quantity $R_H$, which is essentially a ratio of an effective quantum potential to the  elastic potential, we study the behavior of the total geometric  potential as $R_H$ is varied. In the absence of an electron, the elastic potential is positive  and has a single positive maximum for all conformations. Further,  a binormal  helical ribbon conformation has the lowest  potential, while the normal ribbon has the highest, with those of the intermediate ribbons lying in between these. Intriguingly, when a quantum particle is placed on the elastic ribbon,  above a certain critical value of $R_H$, the presence of the quantum geometric potential  {\em reverses}  this order.  But localized states for the particle are not supported.
 Only above a  second critical value of $R_H$, localized states appear for all conformations. Here, the lowest negative minimum of the total potential is reached for the normal ribbon conformation, with its location near the inner edge  of the ribbon. This implies that  the injection of an electron on {\it any} given conformation of the elastic ribbon will induce a conformational transition to the normal ribbon conformation. In the process, the electron gets pushed from inside the ribbon and localizes at  the inner edge,  resulting in a Hall-like voltage.  Our results will be helpful in designing flexible electronic devices, in  addition to  understanding conformational transitions  in a variety of  two-dimensional nanomaterials including biomaterials such as lipid-layer helical ribbons.


\section {Introduction}
Ever since the discovery of the two dimensional nanomaterial graphene \cite{graphene1}  with its amazing  properties such as  a high Young's modulus, i.e, high stiffness (due to strong covalent bonds between carbon atoms) 
 and a low bending rigidity, i.e., high flexibility (due to its monolayer, single atom thickness), various elastic properties of graphene  have  been attracting attention.
  Graphene also  retains its high electrical conductivity even when it is deformed, making it useful in applications such as flexible electronic devices. This has led to the study of  elastic and electrical properties of various {\em other} monolayer materials \cite{2Dmonolayers} such as boron nitride, several dichalcogenides,  complex oxides, etc., as well as  {\em ribbons} made of these nanomaterials. 
  
  In particular,  {\em helical nanoribbons} have been fabricated  using a variety of  materials  that include
 silicon carbide, silicon oxide, zinc oxide,  graphene, transition metal dichalcogenides such as molybdinum disulphide, tungsten disulphide, etc. \cite{ gao, liu, prevost, fan}.  It is by now well recognized that  the electrical properties of  two-dimensional nanomaterials are  sensitively dependent on the {\em twist} between  two layers of the material.  For example, in bilayer graphene it was experimentally found \cite{herrero} that  electronic behavior can change all the way from insulating to superconducting, depending on the interlayer twist angle. This led to a new field called `twistronics'. Recently, an improved method  applicable to {\em general}  two-dimensional (2D) nanomaterials  has been designed \cite{tang}, by using  a microelectromechanical system (MEMS) capable of tuning both the  twist angle and  the interlayer separation simultaneously. It has been demonstrated with twisted hexagonal boron nitride (h-BN) as an example.  In a different recent  experiment, it has been observed  that by twisting layers of tungsten disulfide to form a 3D spiral, a `twistronic Hall effect' \cite{ji} gets generated (in the absence of a magnetic field). In addition, this effect is reminiscent \cite{agarwal} of  the Coriolis effect \cite{goldstein}  that is associated with the deflection of  a moving particle in the transverse direction, as  observed  in rotating (noninertial) systems. 
 
 In a recent paper \cite{PRB2025},
 we considered a quantum particle confined to two special helical ribbon conformations  of a {\em rigid} (or inflexible) helical ribbon,  viz., the normal and binormal ribbons,  and showed  how a  quantum analog of the Coriolis effect emerges essentially due to the twisted geometry of a ribbon.  If the quantum particle is an electron,  our analysis  led to the generation of a Hall-like effect.  
 
 As mentioned above, 2D nanomaterials are typically elastic. Besides, small-width  ribbons made of such materials are usually more flexible than 2D sheets. As is well known, helical ribbons also appear naturally in  {\em biomaterials} \cite{watson,pauling}  which are flexible and elastic. The quantum behavior of a particle confined to an  {\em elastic}  curved surface has not been studied either experimentally or theoretically so far, providing  sufficient motivation for the present work.  It is thus insightful to ask what happens when a quantum particle like an electron is injected on an {\em elastic}  helical nanoribbon of a given conformation. Due to its elasticity, can the ribbon change its original conformation to another one?  What will happen to the above mentioned Hall-like effect? In this paper,  we answer these questions by analyzing a wide class of  elastic helical ribbon conformations.
  
  


 \section{Methodology and summary of results}
 \label{method}
 
 To describe the elastic energy of a 2D nanomaterial, we use the   model proposed by Yang \cite{yang1,yang2}, which is a modified  Canham-Helfrich \cite{canham,helfrich} model.
  The reason for our choice of this model is twofold. First, as has been shown there, certain topological issues encountered in the Helfrich model get resolved in this model. Second, this model is directly applicable to several  {\em nanomaterials}, as  will be elaborated in the Discussion section.
 
  We  consider a  helical ribbon  labelled by an angle parameter  
$\zeta_{0}$, such that each $\zeta_{0}$ corresponds to a specific conformation of the helical ribbon [See Eq. (\ref{S})]. We  use Yang's model to  express the  elastic energy as a function of  its mean curvature $M$ and Gaussian curvature $K$.
We then derive the expressions for $M$ and $K$ for a helical ribbon with a specific $\zeta_{0}$. These expressions in turn enable us to write down the  local elastic  potential for a helical ribbon.

 Next, we write down the Schr\"odinger equation of a particle on a  (rigid, inflexible) curved surface
  using da Costa's formulation \cite{daCosta}. As has been shown  therein, the curved  geometry of the ribbon induces a purely quantum geometric potential  which involves $M$ and $K$. This is an effective potential inherent to the  {\em curved space}, arising essentially from the kinetic energy of the particle on the curved surface. We derive  this potential  for the surface of a  helical nanoribbon, labelled by a specific $\zeta_{0}$ which denotes its conformation.
  
  It is important to note that da Costa's  formulation does not take into account the material properties of the curved surface. In particular, since  a helical ribbon made of  a 2D nanomaterial is {\em elastic},
   it is clear that  its twisted  or deformed surface also  has an elastic potential energy, which is
   stored in the material due to its deformation or  strain. The local strain leads to an internal stress that
  creates a spatially varying potential energy landscape, resulting in a force field  for a quantum particle (such as an electron)  within the material. Therefore, to  describe  an {\em elastic}  helical nanoribbon,  we must {\em include} the  local elastic potential  that we obtained from the elastic model, while analyzing  the Schr\"odinger equation derived by da Costa.  On doing so, we find that the resulting Schr\"odinger equation acquires a {\em total potential} which consists of  a {\em purely quantum} term, and   an  {\em elastic} term.  Further, since both terms in the total (geometric) potential depend on the twisted geometry of the helical ribbon via $M$ and $K$, an  intricate {\em interplay}  emerges between  the  two terms,  leading to several interesting results.  
  
  The paper is organized as follows. In Sec. 3, the parametric equation for the surface of  a helical ribbon is constructed, using its central helix  with  curvature $\kappa_{0}$ and torsion $\tau_{0}$  as the base curve.  We introduce  an angle parameter  $\zeta_{0}$  ($0 \le \zeta_{0}\le 2\pi$ )  which labels a specific conformation of the helical ribbon. In Sec. 4, we determine the first and second fundamental forms 
  \cite{struik} of  its surface and derive the expressions for the mean curvature  $M$ and Gaussian curvature $K$ for the various conformations. In Sec. 5, we discuss the modified Canham-Helfrich model for the elastic energy proposed by Yang \cite{yang1,yang2}, and write down the expression for the local elastic potential for the elastic helical ribbon. It depends on the bending rigidity of the material and the geometrical parameters of the helical ribbon. In Sec. 6, we use da Costa's formulation \cite{daCosta}  to derive the quantum geometric potential  appearing  in the Schr\"{o}dinger equation of a quantum particle confined to a  {\em rigid} helical nanoribbon. In Sec. 7, we use these results  to find the general expression of the total geometric potential for an {\em elastic}  helical nanoribbon. It is convenient to introduce a dimensionless quantity $R_{H}$ [see Eq. (\ref{RH})], which is essentially  the ratio of  an  effective quantum potential to  the elastic potential. In Sec. 8, we present the plots of the  total geometric potential as a function of the width parameter of the helical ribbon   as $R_H$ increases from  $R_{H}=0$, which corresponds to an elastic ribbon in the absence of an electron. We find that to illustrate our findings,  it is sufficient   to plot the total potential for three distinct conformations, viz. the binormal ribbon, the normal ribbon and an intermediate  conformation, with $\zeta_{0} =\pi/2, \pi$ and $3\pi/4$ respectively.  
 
  We demonstrate that 
  a given elastic helical nanoribbon  can change its conformation to another one  with a lower total geometric potential, due to its elasticity, when a quantum particle such as an electron is injected on it. In the absence of an electron,  when $R_{H}=0$,  the elastic potential is positive  and has a single positive maximum for all conformations. Further,  a binormal  helical ribbon conformation has the lowest  potential, while the normal ribbon has the highest, with those of the intermediate ribbons lying in between these. Intriguingly,
when an electron is injected on the elastic ribbon,  above a certain critical value of $R_H$, the presence of the quantum geometric potential 
  {\em reverses}  this order.  However, there are no localized states for the electron.
We show that above a  second critical value of $R_H$, localized states appear for all conformations. Here, the lowest negative minimum of the total potential is reached for the normal ribbon conformation, with its location near the inner edge  of the ribbon. This implies that  the injection of an electron on any given conformation of the elastic ribbon will induce a conformational transition to the normal ribbon conformation. 
During  the above change of conformation, as the  total potential  gets  lowered, the localized state gets pushed towards the inner edge of the helical ribbon.  Thus   a Hall-like voltage is generated  along the width of the ribbon.  Section 9 presents  a  discussion of  nanomaterials for which our model  can be applied. Our results should be useful in designing flexible electronics and also in understanding  conformational transitions in helical nanoribbons made of  a biomaterial.
  
  \section{Parametric equation for  the surface  of a  helical ribbon}
 \label{ribbon}
We begin with the  parametric equation for a  circular helix given by ${\bf R}(s) = [ R_0 \cos\alpha s, R_0 \sin \alpha s, (P_0/2 \pi) \alpha s]$, where $s$ is its arc length parameter.  Here $R_0$ and $P_0$ are its radius and pitch respectively, and $\alpha = 2 \pi/\sqrt{4 \pi^2 R_0^2 +P_0^2}$. The  unit tangent vector for the helix  is ${\bf t}(s)=d{\bf R}/ds$. Now, for  {\em any} curve  in 3D parametrized by $s$, one can define a  unit  normal vector 
${\bf n}(s) ={\bf t}_{s}/|{\bf t}_{s}|$ and a unit binormal  vector  ${\bf b}(s) = [{\bf t}(s) \times {\bf n}(s)]$. These are perpendicular to each other, and  lie on a plane perpendicular to the tangent ${\bf t}(s)$. The  unit orthonormal vector triad   $[{\bf t}, {\bf n}, {\bf b}]$ for a {\em general} curve  satisfies the well known 
Frenet-Serret equations \cite{struik}
\be
{\bf t}_s = \kappa  {\bf n} \,\,; \,\, {\bf n}_s = -\kappa {\bf t}  + \tau {\bf b}\,\,;\,\, {\bf b}_{s} = -\tau {\bf n},
\label{FrS}
\ee
where the subscript $s$ stands for derivative w.r.t $s$, and
$\kappa = |{\bf t}_{s}|$ and $\tau = [{\bf t}.({\bf t}_{s }\times {\bf t}_{ss})]/ \kappa ^{2}$.
 Using these, the curvature  $\kappa$ and the torsion $\tau$ for the helix ${\bf R}(s)$  are found  to be  constants  given by 
$\kappa= \kappa_0 = \alpha^2 R_0 = 4 \pi^2 R_0/ (4 \pi^2 R_0^2 +P_0^2)$ and $\tau = \tau_{0} = \alpha^2 P_0/2\pi = 2\pi P_0/(4 \pi^2 R_0^2 +P_0^2)$. 
It is instructive to write  the parametric equation for the helix in terms of $\kappa_{0}$ and $\tau_{0}$ as
\be 
 {\bf R}(s) =  (1/\alpha^{2})[ \kappa_0 \cos\alpha s, \kappa_0 \sin \alpha s, \tau_{0}\alpha s],
 \label{helix_s}
 \ee 
 where $\alpha^{2} = (\kappa_{0}^2 +\tau_0^2)$.
Using the above circular helix  as  a base curve, we wish to construct a  class of ribbons whose widths lie  in a direction perpendicular to its local tangent ${\bf t}(s)$.  Clearly, such a `width unit vector'  ${\bf d}$ lies on the $[{\bf n}, {\bf b}] $ plane, and is given by
\be
{\bf d}(s, \zeta_{0}) = {\bf n}(s) \cos \zeta_{0} + {\bf b}(s) \sin \zeta_{0}.
\label{d}
\ee
Here, $0 \le \zeta_{0}\le 2\pi$  is an  angle that arises due to the freedom in the choice of  the  direction of  ${\bf d}$ on the $[{\bf n}, {\bf b}] $ plane.

Using Eq. (\ref{d}),  the  parametric equation for the surface of a helical ribbon is given by the position vector
\be
{\bf X} (s, \xi,\zeta_{0}) = {\bf R}(s) + \xi \, {\bf d}(s,\zeta_{0}).
\label{S}
\ee 
Here, the arc length parameter
$s = [0,L]$, where $L$ is the length of the helical ribbon. The width parameter $\xi = [-d,d]$. As should be clear,  for  given values of $\kappa_{0}$ and $\tau_{0}$ of the central helix, each value of 
 $\zeta_{0}$  in Eq. (\ref{S}) corresponds to  a {\em specific  conformation} of the helical ribbon.
  For example,  $\zeta_{0} = \pi$ and  $\pi/2$,  represent the surfaces of the normal and binormal helical ribbons,  given by  ${\bf X}^{(n)} (s, \xi) = {\bf R}(s) -\xi {\bf n}(s)$ and ${\bf X}^{(b)} (s, \xi) = {\bf R}(s) + \xi {\bf b}(s)$, respectively. 

\section {Geometric parameters for a  helical ribbon}

As is well known \cite{struik}, the geometric parameters for a surface  ${\bf X}(s, \xi)$  parameterized by independent variables $s$ and $\xi$  are given in terms of its first  fundamental form $I = d{\bf X}^{2} =  E \,ds^{2} + 2F\, ds d\xi + G \,d\xi^{2}$ and its second fundamental form $II = -d{\bf X}. d{\bf N} = e\, ds^2 + 2f\, ds d\xi + g\, d\xi^{2}$, where the unit normal ${\bf N}(s,\xi)$ at every point $(s,\xi)$ on  a  surface  ${\bf X}(s, \xi)$ 
is defined as ${\bf N}(s, \xi) = \dfrac {({\bf X}_{s} \times {\bf X}_{\xi})}{ |({\bf X}_{s} \times {\bf X}_{\xi})|}$.
 
Using Frenet-Serret equations (\ref {FrS}) for the helix with $\kappa=\kappa_{0}$ and $\tau =\tau_{0}$, we determine
 the  various geometric parameters for  the class of helical ribbons  given in Eq. (\ref{S}) as a function of $\zeta_{0}$ to be
\be
E (\xi,\zeta_{0}) = {\bf X}_s\cdot{\bf X}_s = [1 - \kappa_{0}\cos \zeta_{0}\,\,\,\xi]^{2}\,\, + \tau^{2}_{0}\, \xi^{2}\,\, ;\,\,\,F = {\bf X}_s\cdot{\bf X}_{\xi}\,=0\,\,;\,\,G ={\bf X}_{\xi}\cdot{\bf X}_{\xi}=1 .
\label{EFGS}
\ee
The surface normal  ${\bf N}$  is  found to be
${\bf N}(\xi,\zeta_{0})=  \big[(1-k_{0}\cos \zeta_{0}\,\,\, \xi)( {\bf b} \cos\zeta_{0} -{\bf n} \sin \zeta_{0}) -\tau_{0} \xi {\bf t}\big]/\sqrt {E(\xi)}$, from which we obtain the coefficients of  II as
\be
e \,= - {\bf X}_s\cdot {\bf N}_s\,= -k_0 \sin \zeta_{0} \sqrt{E(\xi,\zeta_{0})};\,\,\,f\, = - {\bf X}_s\cdot {\bf N}_{\xi}\,= \frac{\tau_{0}}{\sqrt{E(\xi,\zeta_{0})}}
\,\,;\,\, g  =  -{\bf X}_{\xi}\cdot {\bf N}_{\xi} \, =0. 
\label{efgS}
\ee
Gaussian curvature $K$ and  mean curvature $M$ of the helical ribbon defined in \cite{struik} are found to be
\be
K (\xi,\zeta_{0})=\frac{(eg-f^2)}{(EG -F^2)}\,= - \frac{\tau_0^{2}}{E^{2}(\xi,\zeta_{0})} ; ~\,M(\xi,\zeta_{0})= \frac{(gE-2fF+eG)}{2(EG - F^2)} \,=- \frac{k_0 \sin \zeta_{0}}{2 \sqrt{E(\xi,\zeta_{0})}}.
\label{KMS}
\ee

 Both the mean curvature $M$ and the Gaussian  curvature $K$ depend on  the specific conformation of the helical ribbon due to their dependence on $\zeta_{0}$.  As pointed out below Eq. (\ref{S}), the  normal and binormal helical ribbons correspond to $\zeta_{0} =\pi$ and $\pi/2$, respectively. It is interesting to note from Eq. (\ref{KMS}) that the mean curvature $M$ is nonzero  for all helical ribbons {\em except} the normal helical ribbon for which it vanishes, signifying that it is a minimal surface.
 
 It is well known that given  $M$ and $K$, the {\em principal curvatures}  $k_{1}$ and $k_{2}$ of a 2D curved surface  can be found as the roots of the quadratic equation
 $ k^{2} -2 M k + K = 0$. Hence
\be
k_{1}\,= \,[M + \sqrt{M^{2} - K} ]\,\,\,;\,\,\,k_{2}\, =\,[M - \sqrt{M^{2} - K} ].
\label {pc} 
\ee 
  Since for any physical surface (like a helical ribbon),  the principal curvatures  in Eq. (\ref{pc}) must be {\em real},  the condition  $(M^2 -K ) \ge  0$  must be satisfied. 

\section{Elastic energy of a helical ribbon} 
\label{elastic}
Historically, the study of elastic energy of soft/bio-materials began in the context of a lipid layer, by treating it as a two-dimensional elastic surface. 
Given  the principal curvatures $k_{1}$ and $k_{2}$,   the following two models  have been customarily  used to describe the elastic energy to be just the bending energy that leads to the deformation of the lipid layer.  For the  
Canham model \cite{canham}, 
$E_{C}= \frac{1}{2} D_{B}\int_{S} [k^{2}_{1} + k^{2}_{2}]\,\, dS$ and  for the Helfrich model \cite{helfrich}, 
$E_{H}= \frac{1}{2} D_{B}\int_{S} [k_{1} + k_{2} - c_{0}]^{2}\,\, dS$, where the integration is over a general curved surface $S$. In these models, $D_{B}$ is the bending rigidity (which is assumed to be the same for both principal curvatures)  and $c_{0}$ is the spontaneous curvature which takes into account the  natural bending of the surface present in the absence of external forces. 
 
More recently,  Yang \cite{yang1,yang2} has presented a detailed theoretical analysis to show that the presence of the spontaneous curvature in  the Helfrich model  obstructs the existence of an energy minimizer for some special closed surfaces. Further, there are no topological bounds for the Helfrich energy.   To remedy these issues, an anisotropic generalization of the Canham  model \cite{canham}   has been suggested in \cite{yang1}. This anisotropic, scale invariant model  for closed surfaces is given by
\be
 E_{Y} = \frac{1}{2} \int_{S} [D_{1} k^{2}_{1} + D_{2} k^{2}_{2}]\, dS.
 \label{EYang}
 \ee
  Here, the two distinct bending rigidities $D_{1}, D_{2} >0$  take into account the possible anisotropy of the  two-dimensional nanosurface. 
  Substituting the expressions  for $k_{1}$ and $k_{2}$ for a general surface  in terms of its mean curvature and Gaussian curvature from Eq. (\ref{pc}) into  Eq. (\ref{EYang}), we obtain
\be 
E_{Y} = \frac{1}{2}\int_{S} \Big[(D_{1}+D_{2}) \big(2 M^{2} - K\big ) + |(D_{1}-D_{2})| 2M \sqrt{M^{2} -K}\Big] dS.
\label{EL_S}
\ee 

At this stage, a remark is in order. The determination of  consistent  values of  the bending rigidities $D_{1}$ and $D_{2}$ of monolayer 2D materials is very challenging,  both theoretically and experimentally, 
since their  values  depend sensitively on the different approaches used to find them.  Further, most papers  give the values only for graphene,  whereas  our work  is generally  applicable to all types of 2D nanomaterials.  In Ref. \cite{kumar},
the bending rigidies have been determined theoretically for monolayer surfaces of {\em many types of nanomaterials}, using an efficient  version of  {\em ab initio} density functional theory. 
Additionally, from Table 1 of  Ref. \cite{kumar}, we find that there exist several materials whose $D_{1}$ and $D_{2}$  values are almost equal.  Hence  the second term in Eq. (\ref{EL_S}) 
can be neglected for such nanomaterials. 
In the Discussion section, we give specific examples of materials  with $D_{1} \approx D_{2}$. For such materials, Eq. (\ref{EL_S}) becomes
\be 
E_{elastic} = D_{1} \int_{S} \big(2 M^{2} - K \big )  dS.
\label{Eelastic}
\ee 
For a helical ribbon  surface given in Eq. (\ref{S}), Eq. (\ref{Eelastic}) therefore reads
\bea
E_{elastic}(\zeta_{0})  &=& D_{1}\,\int_{0}^{L}ds \int_{-d}^{d}\, d\xi \,\sqrt{E(\xi,\zeta_{0})}\, [ 2\,M^{2}(\xi,\zeta_{0})  -\, K(\xi,\zeta_{0})] \nonumber \\ 
&=& LD_{1}  \int_{-d}^{d}\,d\xi \,\,\sqrt{E(\xi,\zeta_{0})}\,\,\, [ 2\,M^{2}(\xi,\zeta_{0})  -\, K(\xi,\zeta_{0})].
\label{Eelastic-general}
\eea
It is convenient to transform to the  dimensionless variable $\bar{\xi} =\xi/d$,  yielding
\bea
E_{elastic}(\zeta_{0}) &=& L\,\,d\,D_{1} \int_{-1}^{1}\,d {\bar{\xi }}\,\,\sqrt{E(\bar{\xi},\zeta_{0})} \,\,\, [ 2\,M^{2}(\bar{\xi},\zeta_{0})  -\, K(\bar{\xi},\zeta_{0})] \nonumber \\ 
&=&  L\,\,d\,D_{1} \int_{-1}^{1}\,d {\bar{\xi }}\,U_{elastic}(\bar{\xi},\zeta_{0}) .
\label{EelasticBar}
\eea
Substituting for $M$ and $K$ from Eq. (\ref{KMS}),  we find $(2M^{2} -K) = \Bigg[\dfrac{\kappa^{2}_{0}\, \sin^{2} \zeta_{0} } {2 E (\bar{\xi,} \zeta_{0})} 
+ \dfrac{\tau^{2}_{0}}{E^{2}(\bar{\xi},\zeta_{0})}\Bigg]$. Using this in Eq. (\ref{EelasticBar}), $U_{elastic}(\bar{\xi},\zeta_{0})$, the {\em local elastic potential at $\bar{\xi}$}  on the surface of a helical ribbon is found to be
\be
U_{elastic}(\bar{\xi},\zeta_{0})  = L\,d\,D_{1} \,\,\left[\dfrac{\kappa^{2}_{0}\, \sin^{2} \zeta_{0} } {2 E^{\frac{1}{2}}(\bar{\xi,} \zeta_{0})} 
+ \dfrac{\tau^{2}_{0}}{E^{\frac{3}{2}}(\bar{\xi},\zeta_{0})}\right],
 \label{UelasticBar}
\ee
where $-1 \le \bar{\xi} \le 1$.
As seen from  Eq. (\ref{EFGS}) , $E(\bar{\xi}, \zeta_{0})$  is always positive.
Since $D_{1} >0$,  we find that the elastic potential $U_{elastic}(\bar{\xi},\zeta_{0})$  is positive definite.


A brief analysis shows that Eq. (\ref {UelasticBar})
  has a {\em single maximum}  at
\be
\bar{\xi}_{max} =  \frac{ \kappa_{0}\,\cos \zeta_{0}}{d\,\,[\kappa^{2}_{0} \cos^{2}\,\zeta_{0} +  \tau^{2}_{0}]}.
\label{xiMaxElast}
\ee
Note that as $\zeta_{0}$ is increased continuously from $\pi/2$ to $\pi$,
  this single maximum value obtained moves continuously from the {\it center} of the ribbon  which corresponds to ($\xi= \bar{\xi} =0$),  towards the inner edge of the helical ribbon, since  $\xi = \bar{\xi}\,d = -\,\dfrac{\kappa_{0}}{[\kappa^{2}_{0} +  \tau^{2}_{0}]} $. 
  
  From Eq. (\ref{xiMaxElast}), we see that  the location of the maximum for a given 
  $\zeta_{0}$ is  dependent on 
  $\kappa_{0}$ and $\tau_{0}$ of the helical ribbon. So by {\em tuning}  these parameters, we can shift the position where the maximum of the elastic potential (\ref {UelasticBar}) occurs.

\section{Quantum geometric potential  on a  helical  nanoribbon}
\label{quantum}
 To investigate the quantum energy associated with a particle constrained to move on  a   helical ribbon defined in Eq. (\ref{S}), we  first write down  the Schr\"odinger equation for the particle on its surface.  A consistent formulation valid for a  quantum particle on a {\em general}  two-dimensional curved surface 
   ${\bf X}(s,\xi)$ embedded in three dimensional  space has been presented by da Costa \cite{daCosta}.
 A lengthy and detailed analysis yields the time-independent Schr\"odinger equation for the surface wave function $\chi(s,\xi)$, valid for all surfaces with orthogonal 
 curvilinear coordinates, i.e., $F={\bf X}_s.{\bf X}_\xi =0$ in the  surface metric. It is given by 
 $-\dfrac{\hbar^2}{2m}\dfrac{1}{h_{1} h_{2}} \Big [\dfrac{\partial}{\partial s}\dfrac{h_{2}}{h_{1} }\dfrac{\partial \chi}{\partial s}  + \dfrac{\partial}{\partial \xi}\dfrac{h_{1}}{h_{2} }\dfrac{\partial \chi}{\partial \xi}\Big]-\dfrac{\hbar^2}{2m}\big[M^2- K\big]\chi = \mathcal{E}\,\,\chi$,
 where  $h_{1}$, $h_{2}$ are the Lam\'{e}  coefficients. Here, $M$ and $K$ are the mean curvature and Gaussian curvature for the general surface. If the surface is {\em elastic}, an elastic potential gets added to the left hand side of this equation.
 
 For  a  helical ribbon,  the  surface metric is given by $(d{\bf X})^2= h_{1}^2\, ds^{2 } + h_{2}^{2}\, d\xi^{2}=  E(\xi,\zeta_{0}) ds^{2} + d\xi^{2}$, yielding  $h_{1}= \sqrt {E(\xi,\zeta_{0})}$ and $h_{2}= 1$. Expressing the above Schr\"odinger equation in terms of $E(\xi,\zeta_{0})$, we get 
 \be
 -\dfrac{\hbar^2}{2m}\dfrac{1}{\sqrt{E}} \Big [\dfrac{\partial}{\partial s}\dfrac{1}{\sqrt{E} }\dfrac{\partial \chi}{\partial s}  + \dfrac{\partial}{\partial \xi} \sqrt{E}\dfrac{\partial \chi}{\partial \xi}\Big]-\dfrac{\hbar^2}{2m}\,\,(M^2- K)\,\chi\,
 = \mathcal{E} \,\chi,
 \label{se_E}
 \ee
where  $E(\xi,\zeta_{0})$ is given in  Eq. (\ref{EFGS}).
From   Eq. (\ref{KMS}), the expression for $M^{2}-K$ can be  found. The Schr\"odinger equation, Eq. (\ref{se_E}), is valid for {\em all} {\em elastic} helical ribbons classified by $\zeta_{0}$. 

Now, the  normal and binormal ribbons  correspond to $\zeta_{0}=\pi$ and $\pi/2$, respectively. The  analysis of Eq. (\ref{se_E}) for these two {\em special cases} were presented individually in our recent paper \cite{PRB2025}.  Further, the ribbons were considered to be {\em rigid}. The analysis  for a {\em general } (rigid) helical ribbon, which has a nontrivial dependence on $\zeta_{0}$  follows essentially the same steps used in that paper. In what follows, we outline only the salient steps  of the lengthy calculation involved, so that this paper is self-contained and comprehensive. 
  
The curved surface area element of the general ribbon is $da =\sqrt{E(\xi,\zeta_{0}))} ds d\xi$.  We first define
$\chi (s,\xi) = \dfrac{\Phi(s,\xi)}{[E(\xi ,\zeta_{0})]^{1/4}}$, so that $\int\int |\Phi|^{2} ds\,\, d\xi =1$. Using this definition, the Schr\"odinger equation  in terms of the  normalized wave function $\Phi(s,\xi)$  is written down. 
We  then look for a separable solution of the form
$\Phi(s, \xi) = u(s) w(\xi)$, and  obtain 
\be
-\dfrac{\hbar^2}{2m}  \dfrac{\partial ^{2} u}{\partial s^{2}} = \mathcal{E}_{0}\, u(s) \,,
\label{u}
\ee
and
\be
-\, \dfrac{\hbar^2}{2m}  \dfrac{\partial ^{2} w(\xi)}{\partial {\xi}^{2}} + \left[V_{eff}(\xi,\zeta_{0})  + \dfrac{\mathcal{E}_{0}} {E(\xi,\zeta_{0})}\right] w(\xi) = \mathcal{E} w(\xi),
\label{se_w}
\ee
where $V_{eff} (\xi,\zeta_{0}) =  - \dfrac{\hbar^2}{2m}  \left[\dfrac{3 (E_{\xi})^{2}}{16 E^{2}(\xi,\zeta_{0}))} -\dfrac{E_{{\xi}{\xi}}} {4 {E(\xi,\zeta_{0})}}  + (M^2- K)\right ]$. Here,  $(M^2- K)$ depends on $\xi$ and $\zeta_{0}$, as can be seen from Eq. (\ref{KMS}). Further, $E_{\xi}$ and $E_{{\xi}{\xi}}$ represent the  first and second derivatives of $E(\xi,\zeta_{0})$ with respect to $\xi$, respectively. 
 The quantity $\mathcal{E}_{0}$ 
 is found by solving Eq. (\ref{u}) for
 the wave function $u(s)$.  Since the particle is confined  to a helical ribbon of length $L$,  the wave function $u(s)$  must vanish at the two ends of the ribbon, giving the boundary condition $u(0)= u(L) =0$. 
 For the solution
 $u(s)= u_{0} \sin {\rm {k}} s $,  and for a
 helical ribbon with length   $L=L (q) =  2 \pi q/ \sqrt{(\kappa_{0}^2 +\tau_0^2)}$  which is composed of integer $q$  complete  $2 \pi$  turns of the helix, we get
 \be
 \rm {k}={\rm k}_{n,\,q} = n\pi /L(q)= (n/2q) \sqrt{(\kappa_{0}^2 +\tau_0^2)},
 \label{knq}
 \ee
 where   $n = 1, 2, 3, ...$ and $q=1,2,3, ...$.  Hence ${\rm k}_{n,\,q}$ is always nonzero. Thus  Eq. (\ref{u}) 
  yields $\mathcal{E}_{0} = \dfrac{\hbar^2 {\rm k} ^{2}}{2m} = \dfrac{\hbar^2 {\rm k}^{2}_{n,q} }{2m}$ to be nonzero as well.  Substituting this in Eq. (\ref {se_w}), we get 
 \be
-\, \dfrac{\hbar^2}{2m}  \dfrac {\partial ^{2} w(\xi)}{\partial {\xi}^{2}} + U_{Q} (\xi, \zeta_{0},{\rm k_{n,q}}) w(\xi) = \mathcal{E} w(\xi) \,, 
\label{se_UQ}
\ee
where 
\be
U_{Q} (\xi, \zeta_{0},{\rm k_{n,q}})
 =  - \dfrac{\hbar^2}{2m} \left[\dfrac{3 (E_{\xi})^{2}}{16 E^{2}(\xi,\zeta_{0}))}
   -\dfrac{E_{{\xi}{\xi}}} {4 {E(\xi,\zeta_{0})}}  + \left(\frac{\kappa_{0}^{2} \sin^{2}\zeta_{0}}{4 E(\xi,\zeta_{0}) } +  \frac{\tau_{0}^{2}}{E^{2}(\xi,\zeta_{0})} \right) - \dfrac{{\rm k}^{2}_{n,q} }{E(\xi,\zeta_{0})}  \right ] . 
\label{UQ}
\ee
Equation (\ref{UQ}) shows that  the geometry induced potential $U_{Q}$ appearing
in Eq. (\ref{se_UQ}) is  purely {\em quantum} mechanical in origin, due to the $\hbar$ dependent prefactor.  Hence $U_{Q}$ is called  the {\em quantum geometric potential}. Here,  $E(\xi, \zeta_{0})$ is defined in Eq. (\ref{EFGS}).
Substituting  its first and second derivatives  $E_{\xi}$ and $E_{\xi \xi}$ 
  together with  $\rm{k}^{2}_{n,q}$ found  in Eq. (\ref{knq}), a very {\em lengthy calculation} reduces the quantum geometric potential given in Eq. (\ref{UQ}) to the following more compact expression, which is valid for all ribbons labelled by $\zeta_{0}$. 
Thus, Eq. (\ref{se_UQ})   can be written as
 \be
-\, \dfrac{\hbar^2}{2m}  \dfrac {\partial ^{2} w(\xi)}{\partial {\xi}^{2}} + U_{Q} (\xi, \zeta_{0},n,q) w(\xi) = \mathcal{E} w(\xi) \,, 
\label{se_UQnq}
\ee
where $U_{Q} (\xi, \zeta_{0},n,q)$ is given by
\be
U_{Q}  (\xi,\zeta_{0}, n,q) = - \dfrac{\hbar^2}{8m} \left[\dfrac{\,(\kappa_0^{2}+\tau_{0}^{2})\left( 1 - \frac{n^{2}}{q^{2}}\right)}{E(\xi,\zeta_{0})} + \dfrac{\tau_0^{2}}{ E^{2}({\xi},\zeta_{0})} \right] .
\label{UQnq}
\ee
The particle quantum number is $n=1,2, 3,...,q$ and the number of full turns of the ribbon is  $q=1,2,3,...$. 

As we did in the discussion of the elastic energy expression in Sec. \ref{elastic}, here too we define a dimensionless variable 
$\bar{\xi} =\xi/d$,  so that  Eq. (\ref{se_UQnq}) becomes
\be
-\, \dfrac{\hbar^2}{2m d^{2}}  \dfrac {\partial ^{2} w(\bar{\xi)}}{\partial {\bar{\xi}}^{2}} + U_{Q} (\bar{\xi}, \zeta_{0},n,q) w(\bar{\xi}) = \mathcal{\bar{E}} w(\bar{\xi}) \,, 
\label{se_UQBar}
\ee
where
\be
U_{Q}  (\bar{\xi}, \zeta_{0}, n,q) =  - \dfrac{\hbar^2\, (\kappa_0^{2}+\tau_{0}^{2}) }{8m}  \,\,  \left[\dfrac{\,\left( 1 - \frac{n^{2}}{q^{2}}\right)}{E(\bar{\xi},\zeta_{0})} + \dfrac{\tau_0^{2}/(\kappa_0^{2}+\tau_{0}^{2})}{E^{2}({\bar{\xi},\zeta_{0}})} \right].
\label{UQnqBar}
\ee
 
A brief analysis shows that the potential 
$U_{Q}(\bar{\xi},\zeta_{0}, n,q)$  for a helical ribbon labelled by $\zeta_{0}$ given in Eq. (\ref{UQnqBar})  has a  single (negative) minimum  at 
\be
\bar{\xi}_{min} =  \frac{ \kappa_{0}\,\cos \zeta_{0}}{d\,\,[\kappa^{2}_{0} \cos^{2}\,\zeta_{0} +  \tau^{2}_{0}]}.
\label{xiMinQuant}
\ee

It is indeed intriguing  that  for all values of $\zeta_{0}$,  the single {\em minimum}  of the {\em quantum}  geometric potential in Eq. (\ref{xiMinQuant}) appears at the {\em same location on the ribbon width} as the single {\em maximum}  of the {\em elastic}  potential that we had found in Eq. (\ref{xiMaxElast}). 
 Further, while the minimum of the quantum potential moves from the center of the ribbon towards the inner edge of the helical ribbon, as 
$\zeta_{0}$ is increased from $\pi/2$ to $\pi$, it is the maximum of the elastic potential that does so.  These behaviors suggest that there is an  intricate interplay between quantum mechanics of a particle on curved nanomaterials, and the elastic property of the nanomaterial. 


\section{ Elastic helical  nanoribbon: Total geometric potential}
\label{total}

We find the Schr\"odinger  equation for an {\em elastic} helical nanoribbon by  adding
 the elastic potential given in Eq. (\ref {UelasticBar}) to Eq. (\ref{se_UQBar}) to yield
 \be
-\, \dfrac{\hbar^2}{2m d^{2}}  \dfrac {\partial ^{2} W(\bar{\xi},\zeta_{0})}{\partial {\bar{\xi}}^{2}} + U_{total} (\bar{\xi}, \zeta_{0},n,q) W(\bar{\xi},\zeta_{0}) = \mathcal{\bar{E}'} W(\bar{\xi},\zeta_{0}) \,, 
\label{se_UtotalBar}
\ee
where $W(\bar{\xi},\zeta_{0})$ is the wave function for the general elastic ribbon with energy eigenvalue $\mathcal{\bar{E}'}$. We have
\be
U_{total}  (\bar{\xi}, \zeta_{0},n,q)=  \Big[U_{Q} +U_{elastic}\Big],
\label{Utotal}
\ee 
giving
\be
U_{total}(\bar{\xi}, \zeta_{0},n,q) = - \dfrac{\hbar^2\, (\kappa_0^{2}+\tau_{0}^{2}) }{8m}\,\left[\dfrac{\,\left( 1 - \frac{n^{2}}{q^{2}}\right)}{E(\bar{\xi},\zeta_{0})} + \dfrac{\tau_0^{2}/(\kappa_0^{2}+\tau_{0}^{2})}{E^{2}({\bar{\xi},\zeta_{0}})} \right]
+D_{1}\,L\,d \,\,\left[\dfrac{\kappa^{2}_{0}\, \sin^{2} \zeta_{0} } {2 E^{\frac{1}{2}}(\bar{\xi,} \zeta_{0})} 
+ \dfrac{\tau^{2}_{0}}{E^{\frac{3}{2}}(\bar{\xi},\zeta_{0})}\right], 
\label{UtotalnqD1}
\ee
where $E(\bar{\xi}, \zeta_{0}) = [1 - \kappa_{0}\,\cos \zeta_{0}\,\,\bar{\xi}\,d]^{2}\,\, + \tau^{2}_{0}\, \bar{\xi}^{2}\,d^{2}\,$.
Note that in Eq. (\ref{UtotalnqD1}), $U_{Q} < 0$ for $n < q$ , and  $U_{elastic} > 0$  showing that there is an interplay between the quantum  and elastic contributions. 
Since both terms  in  the total potential depend on the geometric parameters  
$\kappa_{0}, \tau_{0}, \zeta_{0}$ (for  fixed $n$, $q$)   we  call  it the total {\em geometric} potential.

We rewrite Eq. (\ref{UtotalnqD1}) in terms of the width parameter $\xi$ as
\be
U_{total} = D_{1} \,\,L\,d\,(\kappa_0^{2}+\tau_{0}^{2})\,\,\, U(\xi),
\label{UtotalnqD5}
\ee
where
\be
U(\xi) = \,\,\left[\dfrac{\kappa^{2}_{0}\, \sin^{2} \zeta_{0}/(\kappa_0^{2}+\tau_{0}^{2}) } {2 E^{\frac{1}{2}}(\xi ,\zeta_{0})} 
+ \dfrac{\tau^{2}_{0}/(\kappa_0^{2}+\tau_{0}^{2})}{E^{\frac{3}{2}}(\xi,\zeta_{0})}\right]\,  -  R_{H} \,\left[\dfrac{\,\left( 1 - \frac{n^{2}}{q^{2}}\right)}{E(\xi,\zeta_{0})} + \dfrac{\tau_0^{2}/(\kappa_0^{2}+\tau_{0}^{2})}{E^{2}(\xi,\zeta_{0})} \right]\,,
\label{Uxi}
\ee
with
\be
R_{H} = \dfrac{\hbar^2}{8m L d \,D_{1} }= \dfrac{\hbar^2/8m Ld}{ D_{1}}.
\label{RH}
\ee

Note that $R_{H}$ is dimensionless. It can be regarded as a {\em ratio} of an effective quantum  potential  to the elastic potential. In Eq. (\ref{Uxi}), $E(\xi, \zeta_{0})$  is given in the first entry of  Eq. (\ref{EFGS}).

 \section{Conformational transitions in an elastic  helical nanoribbon}
In this section, we study the behavior of the total geometric potential $U_{total}$ of the helical ribbon 
[Eq. (\ref{UtotalnqD5})], 
by plotting $U({\xi})$ [Eq. (\ref{Uxi})]
as a function of the ribbon's width parameter $\xi$ in Figs. 1, 2 and 3.
 To illustrate our results, we  use
   three conformations,   $\zeta_{0} =\pi/2 $ (binormal ribbon), $3\pi/4$  (intermediate ribbon) and $\pi$ (normal ribbon), for each $R_{H}$, with parameters  $\kappa_{0} =\tau_{0}=1$, $n=1$  and  $q =5$.
   As we have seen,  the value of $\xi _{m} =\frac{ \kappa_{0}\,\cos \zeta_{0}}{\,\,[\kappa^{2}_{0} \cos^{2}\,\zeta_{0} +  \tau^{2}_{0}]}$ corresponds to a {\em minimum} for the quantum geometric potential, and a {\em maximum} for the elastic potential. 
   
Figure 1 gives the plots of $U(\xi)$ as a function of $\xi$, as $R_{H}$   increases from zero to $0.3$.  Note that $R_H = 0$ corresponds to an elastic ribbon {\em without an electron on it}. Here, the binormal ribbon has the lowest (positive) maximum for the total potential, and the  normal ribbon the highest. This behavior continues when an electron is injected, only  till  $R_H$  reaches  a critical value of $0.3$,  when it gets totally  {\em reversed}, with the binormal ribbon attaining the highest maximum and the normal ribbon the lowest. Since the potentials have single positive maxima till this value, localized states for the electron are not supported.  
 \begin{figure}[t]
	\begin{center}
		\includegraphics[scale=0.3]{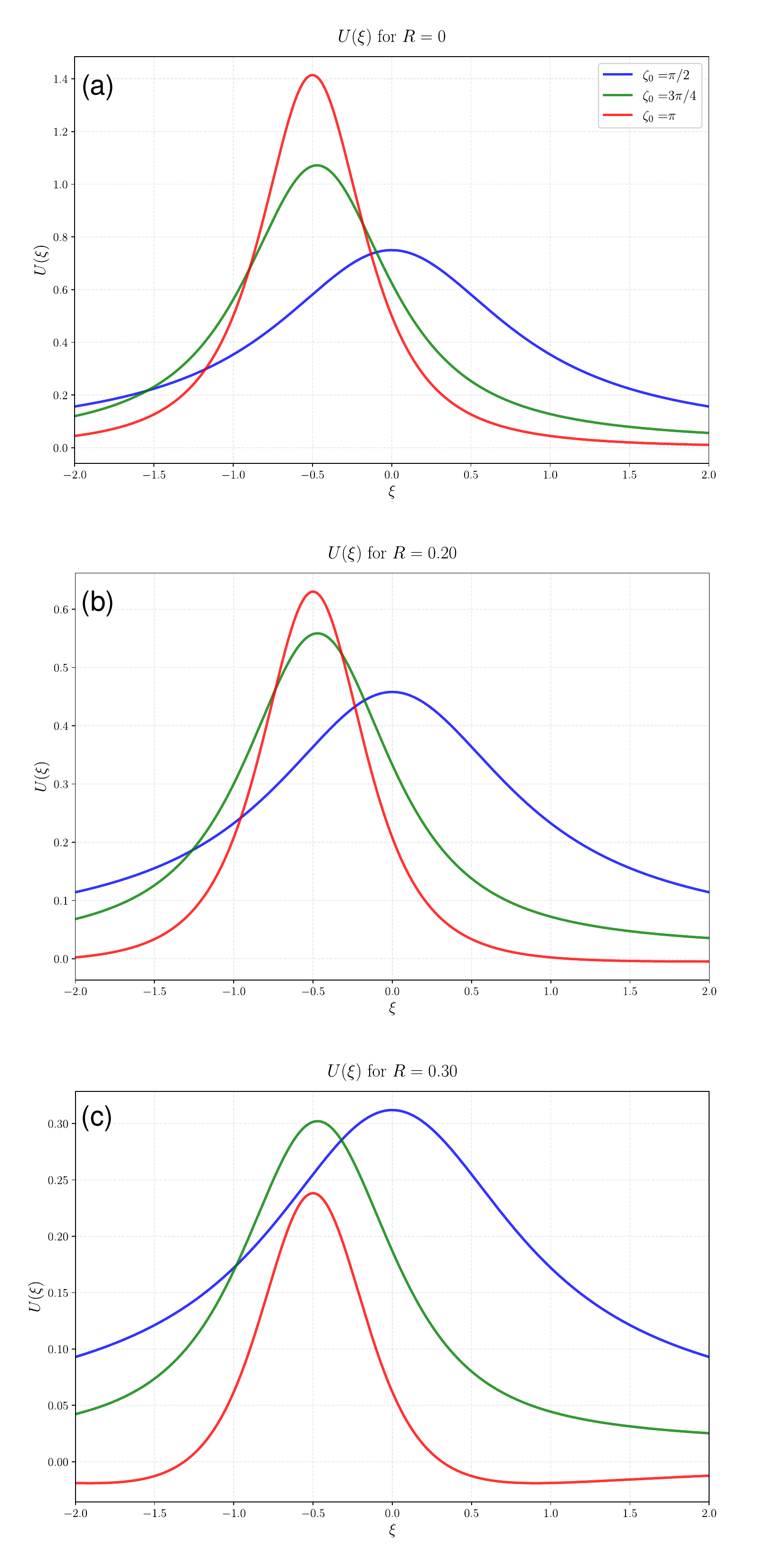}
		\caption{\label{fig:1}  Plots of the total potential $U(\xi)$ [Eq. (\ref{Uxi})] for 
		(a) $R_H=0$, (b) $R_H=0.2$ and (c) $R_H=0.3$. For each $R_H$,  three conformations are plotted: binormal  ribbon ($\zeta_{0} =\pi/2$, blue curve) an intermediate ribbon  ($\zeta_{0} =3\pi/4$, green  curve) and a normal helical ribbon ($\zeta_{0} =\pi$, red curve). $R_H=0$ corresponds to the absence of an electron. Here, the binormal ribbon conformation has the lowest (positive) maximum for the total potential, and the  normal ribbon  has the highest. As the  critical value $R_H=0.3$ is reached, the presence of an electron leads to a total reversal of potentials, with the binormal ribbon having the highest maximum
		and the normal ribbon the lowest. 
		}
	\end{center}
\end{figure}

In Fig. 2, we  plot $U(\xi)$,  as $R_H$  increases from  $0.35$  to $0.45$. The  potentials for the three conformations undergo nontrivial changes in their functional dependence and their extrema, with   positive maxima of some conformations transforming to minima. This is due to the intricate  interplay between the elastic and quantum geometric potential terms  in the total potential.

\begin{figure}[t]
	\begin{center}
		\includegraphics[scale=0.3]{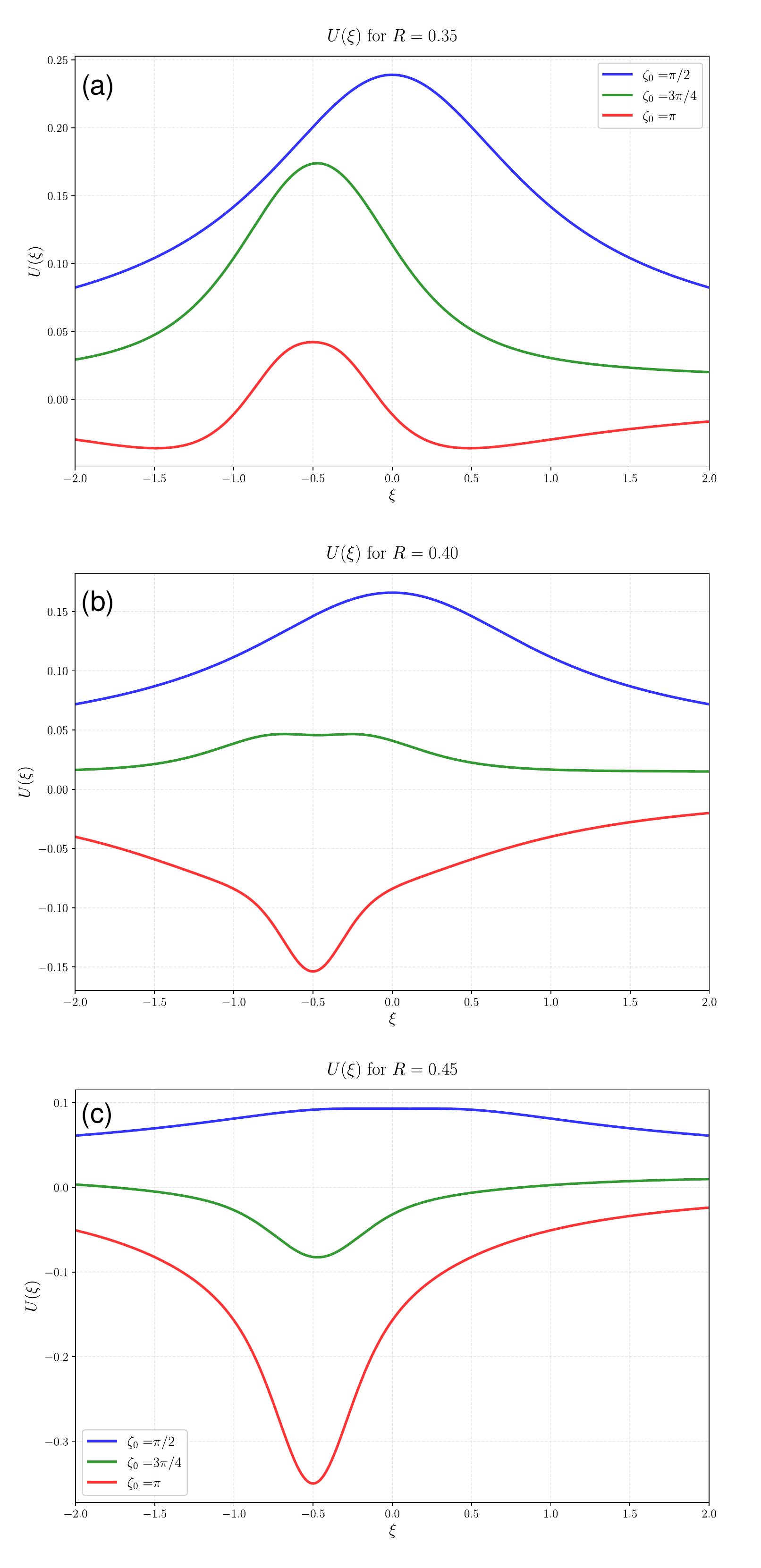}
		\caption{\label{fig:1}  Plots of the total potential $U(\xi)$ [Eq. (\ref{Uxi})] for (a) $R_H=0.35$, (b) $R_H=0.4$ and (c) $R_H = 0.45$. For each $R_H$,  three conformations are plotted: binormal  ribbon ($\zeta_{0} =\pi/2$, blue curve), an intermediate ribbon  ($\zeta_{0} =3\pi/4$, green  curve) and a normal helical ribbon ($\zeta_{0} =\pi$, red curve). Note the nontrivial changes in the plots as $R_H$  gradually increases in this range.		}
	\end{center}
\end{figure} 

Figure 3  gives the plots of $U(\xi)$ as  $R_{H}$  increases from $0.5$ to $2.0$.  As a (second) critical value $R_{H}=0.5$ is reached,  {\em all } the conformations attain single  negative minima, showing that  localized states are supported for all of them. Here, the binormal ribbon has the lowest negative single minimum for the total potential, and the  normal  ribbon has the highest (deepest) negative minimum. In these cases,  we  see from each plot  that as $\zeta_{0}$ for a helical ribbon is increased continuously from $\zeta_{0} =\pi/2$  (for the binormal helical ribbon) to $\zeta_{0} =\pi$
(for the normal ribbon)  the  minimum  value of the  total geometric potential   decreases continuously. Simultaneously, for each $R_H$ value, the position  of the minimum where the electron localizes, moves continuously from the center of the ribbon ($\xi_{min} =0$) towards the inner edge of the ribbon ($\xi_{m} = -\frac{ \kappa_{0}}{[\kappa^{2}_{0} +  \tau^{2}_{0}]}$). 
This leads us to conclude that  for any $R_H$ value in this range, when an electron is injected on any conformation of the elastic helical ribbon, it will undergo a conformational transformation to the normal ribbon, since it has the lowest total potential. While lowering of the total potential of the system is taking place, the electron gets pushed and  localizes at the inner edge of the helical ribbon, signalling a Hall-like effect. Note, however, that there is no magnetic field and this effect arises  due to the curved geometry of the elastic helical ribbon.
\begin{figure}[t]
	\begin{center}
		\includegraphics[scale=0.3]{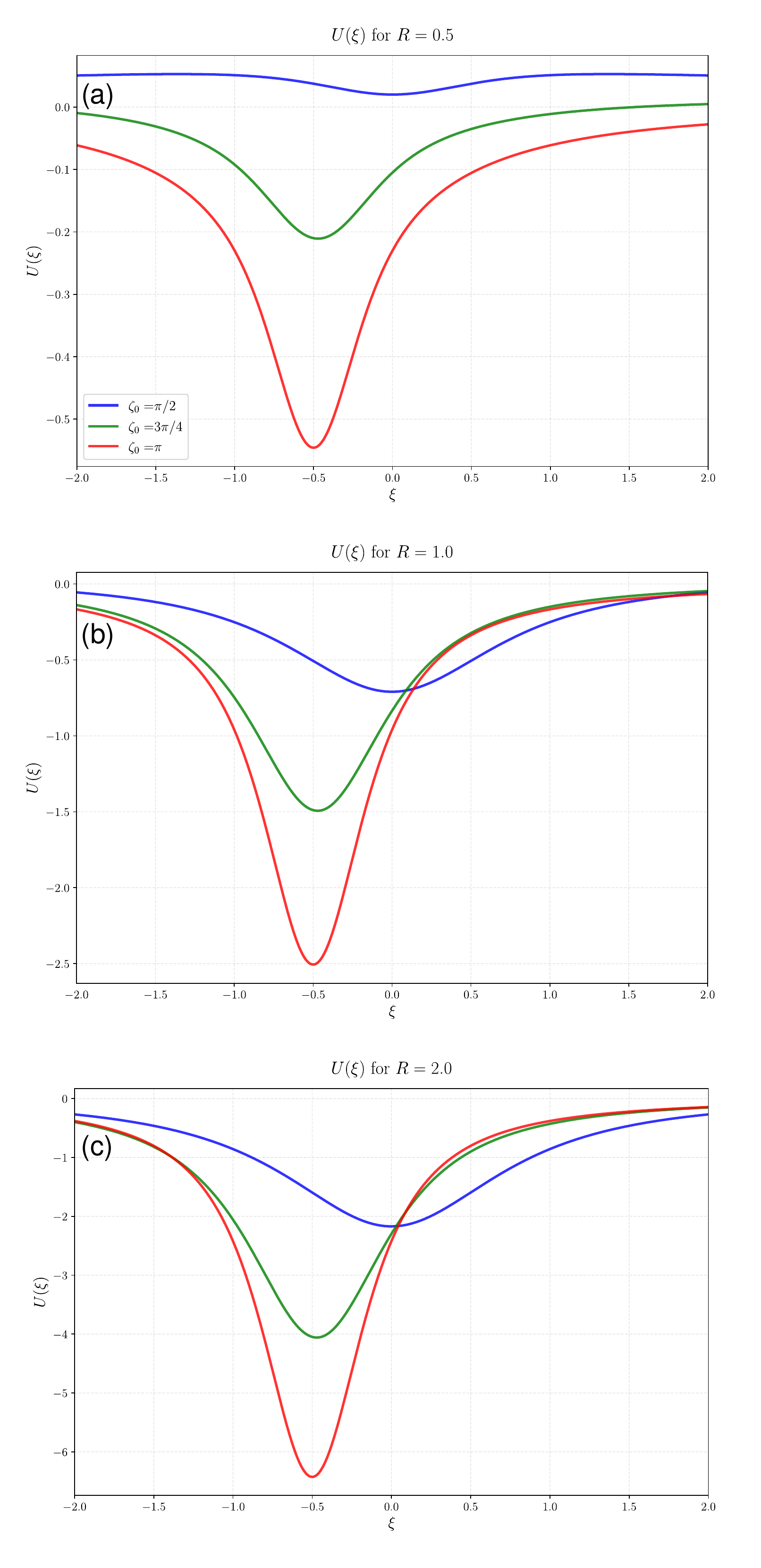}
		\caption{\label{fig:1}  Plots of the total potential $U(\xi)$ [Eq. (\ref{Uxi})] for (a) $R_H=0.5$, (b) $R_H=1.0$ and (c) $R_H = 2.0$. For each $R_H$,  three conformations are plotted: binormal  ribbon ($\zeta_{0} =\pi/2$, blue curve) an intermediate ribbon  ($\zeta_{0} =3\pi/4$, green  curve) and a normal helical ribbon ($\zeta_{0} =\pi$, red curve). The normal ribbon has the lowest (negative) single minimum of the total potential, and the  binormal ribbon highest, with that for the intermediate ribbon lying between the two. The minimum is at the center for the binormal ribbon, and near the inner edge for the normal ribbon. 
}
	\end{center}
\end{figure}

 \section{Discussion}
We have studied conformational transitions in elastic helical ribbons made of  nanomaterials, where the bending rigidity parameters are such that  $|D_{1} - D_{2}|$ is negligible as compared to $2D_{1}$ in Eq. (\ref{EL_S}). However, our formulation can be easily extended to include the term involving $|D_{1} - D_{2}|$. 

As already pointed out in Sec. 5, the determination of  consistent  values of  the bending rigidities $D_{1}$ and $D_{2}$ of monolayer 2D materials is very challenging,  both theoretically and experimentally. Focusing on theoretical results, an inspection of Table 1 in \cite{kumar}, shows that there are several nanomaterials   like C (graphene), Sn, Ge, Si, BN for  which  $D_{1}$ and $D_{2}$ values do not differ much. Further, they are of the order of 1 eV for most of  these materials. (For graphene, 
 $D_{1}$ and $D_{2}$ values are $1.49$ eV and $1.50$ eV, respectively.)  
 
 We make the following interesting observation. The plot for $R_{H} =0$ corresponds to an elastic helical ribbon, with no electrons on it. Here, the binormal helical ribbon has the lowest elastic potential. Therefore, any other conformation (including the normal ribbon)  when perturbed (mechanically, for instance) will 
 attain the binormal ribbon conformation. Our conclusion is consistent with the result  based on the analysis of dynamics  of various conformations of an elastic helical  strip carried out in \cite{goriely}, where it was shown that  a binormal helical strip (ribbon)  is more stable than a normal helical strip.
 
 It is instructive to {\em estimate}  the value of $R_{H}$  [Eq. (\ref{RH})] for some  nanomaterials. In this expression, taking $L$ and $d$ to be in nanometers and $m$ to be the electron mass, we find
${\hbar^2/8m Ld} \approx 0.01$ eV. This yields  $R_{H} =0.01$ for $D_{1} \approx$ 1eV.  Our  plots suggest that localized states will  appear only for $R_{H} =0.5$.  
Interestingly, there exist superflexible materials made of graphene oxide \cite{superflexible} for which $D_1 \approx 1K_{B}T= 0.025$ eV at room temperature.  For these, $R_{H}$  $\approx$ $0.5$. If helical ribbons can be fabricated for these materials, the Hall-like effect can be observed, leading to applications like flexible electronics.
{\em Biomaterials} such as actin and peptides typically possess  very low  $D_{1}$ values. Here, our results show that  conformational changes observed in these biomaterials for various $R_{H}$ values  can perhaps be explained by an electron getting injected on them due to a biochemical process. 
 
 Note that Figs. 1 to 3  obtained for various $R_{H}$ values  have been plotted  by taking $\kappa_{0} =\tau_{0}=1$, $n=1$  and  $q =5$  in the expression for the total potential given in Eq. (\ref{Uxi}). As is clear, by varying these parameters, it should be possible to {\em lower}  the critical value of $R_{H}$ when localized states appear, so that the Hall-like effect can emerge  for helical ribbons made of graphene and other 2D nanomaterials \cite{kumar} with $D_{1}$ of the order of 1 eV.

Our analysis is also applicable  when an electron gets added in a natural fashion to lipid layer helical ribbons  that can appear in biomaterials such as cholesterol \cite{cholesterol}. Such ribbons can form via molecular self-assembly, where lipid molecules arrange themselves in a twisted structure. As we have seen, the presence of  an electron (or ion) will cause  conformational changes which play an important role in biological systems.
Our results should  motivate  the fabrication of helical ribbons made of a  variety of  2D nanomaterials, along with determination of their bending rigidities. It would be interesting to experimentally investigate our prediction of  conformational transitions in helical nanoribbons, as well as the appearance of Hall-like voltages.

\section{Acknowledgments} 
We are indebted to Robin Msiska for the help with the figures.  The work of A.S. at Los Alamos National Laboratory was carried out under the auspices of the U.S. DOE and NNSA under Contract No. 89233218CNA000001.

\end{document}